# Unveiling spatial correlations in biophotonic architecture of transparent insect wings

Pramod Kumar[1], Danish Shamoon[1], Dhirendra P. Singh[1], Sudip Mandal[2] and Kamal P. Singh[1]

[1]*Femtosecond Laser Laboratory, Department of Physical Sciences, Indian Institute of Science Education and Research (IISER) Mohali, Knowledge City, Sector 81, S.A.S. Nagar, Mohali-140306, Punjab, India.*
[2]*Department of Biological Sciences, Indian Institute of Science Education and Research (IISER) Mohali, Knowledge City, Sector 81, S.A.S. Nagar, Mohali-140306, Punjab, India.*
pk6965@gmail.com, kamal@iisermohali.ac.in

Abstract: We probe the natural complex structures in the transparent insect wings by a simple, non-invasive, real time optical technique using both monochromatic and broadband femtosecond lasers. A stable, reproducible and novel diffraction pattern is observed unveiling long range spatial correlations and structural-symmetry at various length scales for a large variety of wings. While matching the sensitivity of SEM for such microstructures, it is highly efficient for extracting long range structural organization with potentially broad applicability.

## 1 INTRODUCTION

Natural Photonics structures (Paris et al., 2012; Mika et al., 2012; Parker, 2009; Xu et al., 2007 ) in the transparent insect wings have attracted much attention in recent years not only because of their potential for various biomimetics technological applications but also as ideal test bed to learn principles of coherent manipulation of light by nature (Wiederhecker et al., 2009; Mathias et. al., 2010; Shevtsovaa et al., 2011). Compared with the equivalent man made optical devices, the biophotonic structures often possess large complexity and even render better performance in some cases (Bar-cohen, 2011). One of the main organizing principles of complex patterns in transparent insect wings is their symmetry and long range correlations at various length scales from nanometer to micrometer structures. Hence the study of symmetry in natural structural arrangements is crucial to explore novel optical effects (Shevtsovaa, 2011; Pouya et al., 2011; Trzeciak et al., 2012). Variations in both dimensionality and degree of periodicity contributes greatly to the over-whelming variety of common optical phenomena reported like reflection, refraction, interference, fluorescence, iridescence, and so forth (Jordan et al., 2012).

Insect wings are a multifunctional material system having various distributions in size and shape, spatial heterogeneity in its structural arrangements, and orientation of the photonic architectures. Despite this quasi-disorder wings are known to manipulate light in a coherent way. Many studies and techniques like SEM/TEM have postulated various explanations for insect wings complexity with high resolution locally. However, a systematic and efficient approach to explore long range structural correlations over the entire length scale of the wing is desired. Because of the absence of any obvious periodicity, such systems are in general difficult to handle in theory and a super-cell is usually needed in numerical simulations (Mihailescu et al., 2012; Kenji Yamamoto et al., 2012). Here we optically probe long range correlations and spatial symmetry of the structural organization of the photonic architecture of wing. Due to the sensitivity of diffraction, our technique matches the local resolution of SEM for such microstructures, yet is highly efficient to extract *in situ* structural organization which would be very tedious otherwise. Understanding these symmetry principles are crucial for biomimetics photonic structures as well as functional significance of the photonic design.

## 2 EXPERIMENTAL SET-UP

The experimental setup primarily consists of a wing sample holder mounted on a precision xy translation stage, a collimated laser beam, and a screen (see Fig. 1 for a real picture). We have used both a monochromatic cw laser and a broadband femtosecond laser in near IR range. The choice of these wavelengths is dictated by the observed transparency of the wing material for these spectral regions. The typical $1/e^2$ beam waist of laser passing through the wing is ~2mm and the diffracted light is collected on a white screen placed about few meters away.

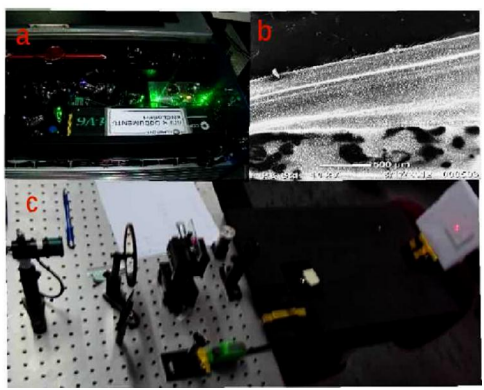

Figure 1: (a) Femtosecond laser system; (b) SEM image of the insect wing surface; (c) Actual picture of the experimental set-up for the diffraction measurement.

The snapshots of the diffraction pattern on the screen is recorded by a digital camera and the obtained images are analyzed using Matlab programming. It is worth mentioning that with this simple setup, no preparation of the wing sample is required and in fact it can be used for *in viva* non-destructive imaging with the insect alive. By scanning the beam spot across the wing various regions can be probed in a single-shot manner.

## 3 RESULTS AND DISCUSSION

A typical diffraction pattern of the wing in transmission is shown for a femtosecond broadband laser cantered at 800nm (top panel in Fig. 2) and with a green 532nm cw laser (bottom panel in Fig. 2). A stable diffraction pattern is clearly visible for both the lasers that exhibit a bright central spot and up to two distinct maxima on either side (see intensity cuts of the corresponding profiles in Fig. 2). The dimensions (1D, 2D, 3D) of periodicity affect the light spectrum which could change the surface colour or forms anti-reflective transparent surface. That is why, we used broadband light pulses to reveal the correlations in structural symmetry at various length scales. Due to complex multi-scale architecture and their local orientational symmetry, the light diffract in such way that the global structures reflect a single homogeneous photonic surface. A SEM imaging of the wing surface (Fig. 3) reveals that it consists of a large quasi-periodic arrangement of micro-hooks on the surface of transparent wing. Clearly, SEM provides local details of the structural arrangement of the shape and short range correlation. However, due to their quasi-periodicity (Liu et al., 2011) these micro-hooks create stable diffraction pattern that is the central observation of this work. Notably, for a mm size beam spot, the pattern is a result of interference created by several thousands of these hooks. In principle, one can extract all possible structural information, including long range distribution, mean spacing between these, and their shape distribution.

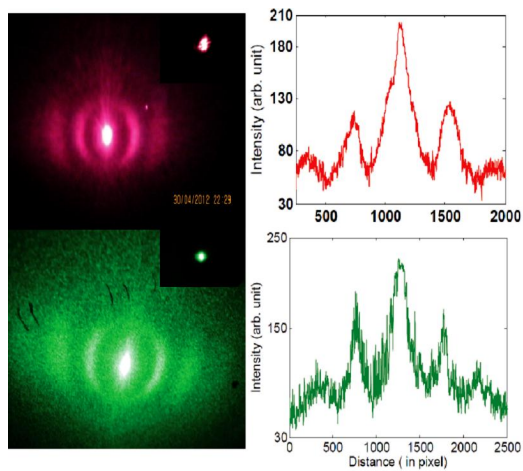

Figure 2: Top left corner: photograph of the diffraction pattern using 7fs, 2nJ@78MHz laser incident beam, the inset in figure of top left corner is the corresponding beam profile; top right corner: ID intensity profile of the diffraction pattern; bottom left corner: photograph of the diffraction pattern using 532 nm green laser beam, the inset in figure of bottom left corner is the corresponding beam profile; bottom right corner: intensity profile of the diffraction pattern.

To analyse the diffraction pattern formation, and hence the structural correlation of photonic structure, on various length scales we need methods that allow two things (i) clear identification of individual surface structures, and (ii) determination of the distribution function of size, shape and the separation of the structures to quantify the structural uniformity, i.e., the degree of organization. The first is of course accomplished by direct imaging, whereas for the second the optical diffraction technique is best suited. To analyse the data one has to distinguish between regular surface structures resulting in diffraction pattern and continuous surface causing a diffuse intensity in the experiment. There are two extreme cases of structural order: the first one is the random arrangement of the nanostructures that is characterised by complete absence of any long range order. These structures show a fractal-like behaviour on the short length scale and is smooth on the long length scale. The other extreme case is a perfectly periodic structured surface that exhibit complete long range order. In this case, the 2D power spectrum consists of a periodic arrangement of sharp peaks as they are known for diffraction grating. From the orientation and symmetry of these peaks the real space orientation and the symmetry of underlying structure can be deduced.

The location of the first order peak in the Fourier space yields real space lateral periodicity of the pattern in the corresponding direction. For a non-identical order, the peaks broaden and the number of detectable higher order peaks decrease. The full width at half maximum of the first order peak is a quantitative measure of the width of the distribution function of the structural separations. In other words, narrower the peaks and larger the number of high-orders, higher is the uniformity of the surface pattern. If the separation between these structures are smaller, i.e., of the order of its size, this produces a more divergent pattern where the distance between zero$^{th}$ and the first order peak is larger. The width of these peaks encodes information about their distribution of shapes and sizes of the individual hook elements. This therefore provides a measure to quantify the organization of such structure. The intricate arrangement of hooks on the surface architectures control how the photons of light interact with them. To prove that this technique, besides being simple and non-invasive, offers sensitivity matching that of SEM, we have taken SEM images of the wing on various scales.

The two-dimensional diffraction pattern (Kenji Yamamoto et al., 2012) by an aperture A(x, y) is the sum of wave produced by the light source at every points of the aperture. The diffracted wave front observed at a distance z from the aperture is given by:

$$A'(x',y') = \frac{z}{i\lambda} \int_{-\infty}^{\infty} \int_{-\infty}^{\infty} A(x,y) \frac{e^{ikr}}{r^2} dxdy \qquad (1)$$

Since we will only observe the intensity of the signal, so it can be shown that the far field diffraction pattern is the Fourier transform of A(x, y). Then the equation (1) can be written as :

$$A'(\nu_x, \nu_y) = FT\{A(x,y)\} \qquad (2)$$

A fast Fourier transform of these images generates a computational far-field diffraction pattern that resembles very closely to the observed pattern (see Fig. 3).

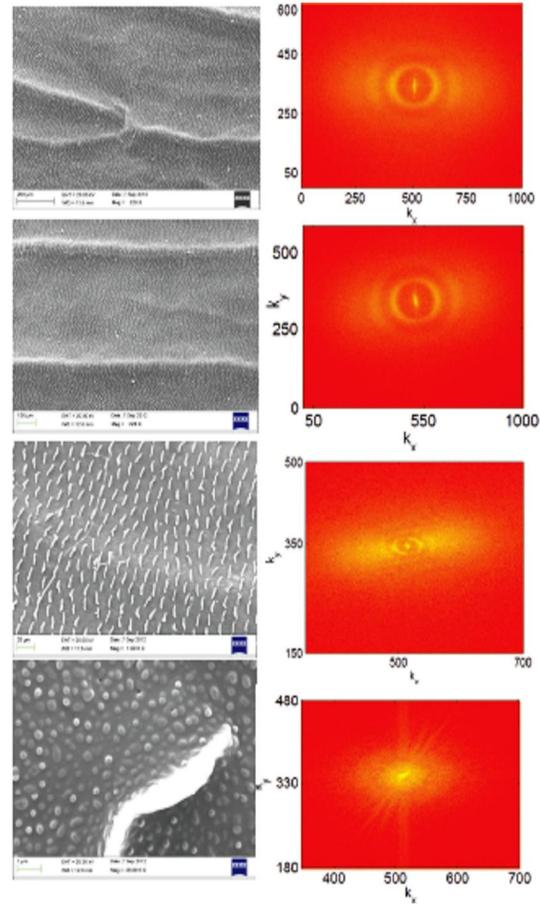

Figure 3: (left column) SEM images of the wing surface at 200µm, 100µm, 10µm, 1µm length scales ; (right column) the right column is the FFT pattern computed for the corresponding images.

The diffraction pattern on various length scales reveal how the higher order peaks and their distribution reflect the structural symmetry and correlation between these hooks. This disorder arrangement will results in broadening of diffraction pattern. As illustrated in Fig 3, the diffraction by a single hook is compared with the corresponding diffraction from an array of hooks at various length scales on the wing surface.

Furthermore, we have observed the rotation of the diffraction pattern when one translates the beam spot across the wing as shown in Fig. 4. This unusual behavior reflects both the local and global patterns in the variation of the distribution symmetry.

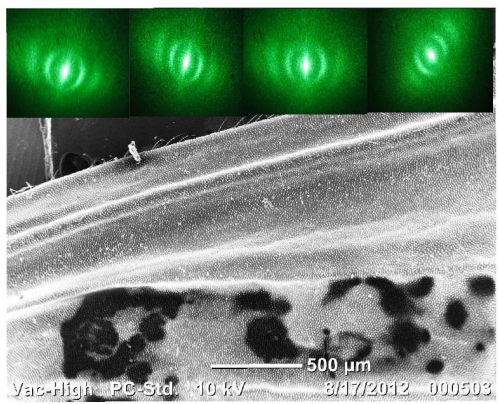

Figure 4: SEM image of the wing, the inset on the top of the image are the rotated diffraction pattern profile at various regions on the wing surface.

It is possible that by varying the spot size from few microns to several mm, one can obtain the spatial correlation in a high throughput, single-shot manner. If such information is attempted by SEM imaging it would be very tedious and inefficient process. The spatial correlation of the light and high sensitivity of the diffraction pattern offer unique advantage over other methods. It is therefore a potentially attractive optical technique to unravel the natural design of the photonic architecture of the insect wing and probe their functional relationship that ultimately dictates the motive of symmetry and correlation.

## 4 CONCLUSIONS

In summary, the proposed optical technique may prove to be a powerful alternative to gain a better understanding about the systematic of photonic architecture such as long range spatial correlations and symmetry in the insect wings. Our finding directly demonstrates how the transmitted diffraction pattern from the wing is correlated with the internal structural symmetry. The rotations of diffraction pattern were obtained when the laser beam scanned various regions on the wing surface as shown in Fig. 4. These rotations of the pattern give the signature of the spatial symmetry in structural arrangements in the local to global length scale. These optical tools could be crucial to understand design principles of natural photonic crystals with potential applications for mimicking artificial structures (Mathias et al., 2010; Bar-cohen, 2011) that may lead to novel optical devices

## ACKNOWLEDGEMENTS

All authors thank to the DST and IISER Mohali, India for supporting this research through grant and research fellowship for Dr. Pramod Kumar. The invaluable help of Babita Basoya and Gopal Verma are grateful acknowledged.